\documentclass[twocolumn,floatfix,superscriptaddress,a4paper,showpacs,showkeys,nofootinbib,reprint,prc]{revtex4-1}
\usepackage[colorlinks=true,linktocpage=true,linkcolor=blue,citecolor=blue,allcolors=blue]{hyperref}
\usepackage{epsfig}
\usepackage{latexsym}
\usepackage{xspace}
\usepackage{threeparttable}
\usepackage{multirow}
\usepackage[utf8]{inputenc}
\usepackage{indentfirst}
\usepackage{enumerate}
\usepackage{color}
\usepackage{tabularx}

\usepackage{amsmath}
\usepackage{amssymb}
\usepackage[english]{babel}
\usepackage{url}
\begin{document}

\title{Application of machine learning in the determination of impact parameter in the $^{132}$Sn+$^{124}$Sn system}

\author{Fupeng Li}
\affiliation{School of Science, Huzhou University, Huzhou 313000, China}
\affiliation{College of Science, Zhejiang University of Technology, Hangzhou 310014, China}
\author{Yongjia Wang}
\email{Corresponding author: wangyongjia@zjhu.edu.cn}
\affiliation{School of Science, Huzhou University, Huzhou 313000, China}
\author{Zepeng Gao}
\affiliation{College of Physics Science and Technology, Shenyang Normal University, Shenyang 110034, China}
\affiliation{School of Science, Huzhou University, Huzhou 313000, China}
\author{Pengcheng Li}
\affiliation{School of Nuclear Science and Technology, Lanzhou University, Lanzhou 730000, China}
\affiliation{School of Science, Huzhou University, Huzhou 313000, China}
\author{Hongliang L\"u}
\affiliation{HiSilicon Research Department, Huawei Technoloies Co., Ltd., Shenzhen 518000, China}
\author{Qingfeng Li}
\email{Corresponding author: liqf@zjhu.edu.cn}
\affiliation{School of Science, Huzhou University, Huzhou 313000, China}
\affiliation{Institute of Modern Physics, Chinese Academy of Sciences, Lanzhou 730000, China}
\author{C. Y. Tsang}
\affiliation {National Superconducting Cyclotron Laboratory and the Department of Physics and Astronomy, Michigan State University, East Lansing, MI 48824, USA}
\author{M. B. Tsang}
\affiliation {National Superconducting Cyclotron Laboratory and the Department of Physics and Astronomy, Michigan State University, East Lansing, MI 48824, USA}

\begin{abstract}
\textbf{Background:} $^{132}$Sn+$^{124}$Sn collisions at the beam energy of 270 MeV$/$nucleon have been performed at the Radioactive Isotope Beam Factory (RIBF) in RIKEN to investigate the nuclear equation of state. Reconstructing impact parameter is one of the important tasks in the experiment as it relates to many observables.\\
\textbf{Purpose:} In this work, we employ three commonly used algorithms in machine learning, the artificial neural network (ANN), the convolutional neural network (CNN) and the light gradient boosting machine (LightGBM), to determine impact parameter by analyzing either the charged particles spectra or several features simulated with events from the ultra-relativistic quantum molecular dynamics (UrQMD) model.\\
\textbf{Method:} To closely imitate experimental data and investigate the generalizability of the trained machine learning algorithms, incompressibility of nuclear equation of state and the in-medium nucleon-nucleon cross sections are varied in the UrQMD model to generate the training data.\\
\textbf{Results:} The mean absolute error $\Delta b$ between the true and the predicted impact parameter is smaller than 0.45 fm if training and testing sets are sampled from the UrQMD model with the same parameter set.
However, if training and testing  sets are sampled with different parameter sets, $\Delta b$ would increase to 0.8 fm.\\
\textbf{Conclusion:} The generalizability of the trained machine learning algorithms suggests that these machine learning algorithms can be used reliably to reconstruct impact parameter in experiment.
\end{abstract}

\maketitle

\section{Introduction}\label{section1}
Heavy-ion collisions (HICs) provide a unique opportunity to explore the nuclear equation of state (EoS), which remains a key requirement for understanding nuclear reaction, nuclear structure, as well as neutron star properties \cite{BALi,Tsang,XuJun,Li:2018lpy,Li:2019xxz,Colonna:2020euy,Ono:2019jxm}. In recent decades, both experimentalists and theorists have made major efforts to obtain information of EoS. These studies reveal that the uncertainty of nuclear EoS is the largest in the density-dependent term at high densities. For this purpose, $^{132}$Sn+$^{124}$Sn, $^{112}$Sn+$^{124}$Sn, and $^{108}$Sn+$^{112}$Sn collisions at beam energy of 270 MeV$/$nucleon have been performed at the Radioactive Isotope Beam Factory (RIBF) in RIKEN \cite{PLBMSU}.

Usually in an experiment, the centrality or impact parameter is reconstructed by using the relationship between observed quantities and the collision geometry.
Recently, the field of artificial intelligence (AI) has received unprecedented attention, and prodigious progress has been made in the application of the AI techniques, see e.g., Ref \cite{Lecun} and references therein. Machine learning (ML), which is a subset of AI, is an interdisciplinary subject, involving probability theory, statistics, approximation theory, convex analysis, algorithm complexity theory and other subjects. Due to the powerful learning and induction ability, ML approaches are widely employed in physical science \cite{PJC,PLB1,PLB2,JEPLB,PLG,SCPMA2020LHL,SCPMA2020BER,ROMP,Nature1,sci1,NP,JAN,DYLEPJC,PRDZHOUKAI,EOS}. Of particular relevance to this work, there are several applications of ML in reconstructing the impact parameter in HICs~\cite{CDPRC,SBASS,Haddad,JD,PLBZK}. For example, in Refs.~\cite{CDPRC,SBASS,Haddad,JD}, the artificial neural network (ANN) or support vector machine is used to reconstruct the impact parameter from final state observables or the particle momentum distributions. In our previous work~\cite{JPGLFP}, we utilized the convolutional neural network (CNN) and the light gradient boosting machine (LightGBM) to determine impact parameter from two-dimensional transverse momentum and rapidity spectra of protons on event-by-event basis. All these aforementioned studies revealed the capability of ML methods in reconstructing impact parameter. We note that in these studies, the training data is usually generated with theoretical models. For example, the quantum molecular dynamics (QMD) model was used to generate data in Refs \cite{CDPRC,SBASS,Haddad}, and a classical molecular dynamics (CMD) model was used in Ref \cite{JD}. When applying these ML methods to analyze real experimental data, the reliability should be evaluated as none of the theoretical models represent real experimental data perfectly. Using data generated from different physical models or different model parameter sets from the same model should give a good estimation of ML's capability. For this purpose, the ultra-relativistic quantum molecular dynamics (UrQMD) model with different nuclear EoS and different in-medium nucleon-nucleon cross sections (two of the main ingredients in the transport model) is used to generate data, and the generalizability of ML methods is investigated by generating training and testing sets with different model parameters.

The paper is organized as follows.
 In Sec.\uppercase\expandafter{\romannumeral2}, we will briefly introduce the UrQMD model and different datasets in this study.
 We continue with Sec.\uppercase\expandafter{\romannumeral3} in which ANN, CNN and LightGBM algorithms will be described.
 In Sec.\uppercase\expandafter{\romannumeral4}, we discuss the results and generalizability of the three algorithms in detail.
 We end with Sec.\uppercase\expandafter{\romannumeral5}, which is dedicated to summary and outlook.
\section{UrQMD Model}\label{section2}
The UrQMD model is a many-body microscopic transport model which has been successfully extended to describe HICs with beam energy from tens of MeV per nucleon up TeV per nucleon available at CERN Large Hadron Collider (LHC) \cite{SAB,BLE,qfli1,qfli2,SCPMA2019QFL,FOP-wyj}. In the UrQMD model, each nucleon is represented by a Gaussian wave packet in phase space.
The coordinates $r_{i}$ and momentum $p_{i}$ of particles \emph{i} are propagated according to Hamilton's equation of motion:
\begin{eqnarray}\label{eq1}
\dot{\textbf{r}}_{i}=\frac{\partial  \langle H  \rangle}{\partial\textbf{ p}_{i}},
\dot{\textbf{p}}_{i}=-\frac{\partial  \langle H \rangle}{\partial \textbf{r}_{i}}.
\end{eqnarray}
Here, {\it $\langle H \rangle$} is the total Hamiltonian function. It consists of the kinetic energy $T$ and the potential energy $U$ with $U=\sum_{i\neq j} V_{ij}$. The following density and momentum dependent potential has been widely employed in QMD-like models, \cite{AIC,qfli3,CH,TLY,YU,feng},
\begin{equation}\label{eq2}
\begin{aligned}
V_{ij}& =\alpha\left(\frac{\rho_{ij}}{\rho_0}\right)+\beta\left(\frac{\rho_{ij}}{\rho_0}\right)^{\eta} \\
       &+ t_{md} \ln^2[1+a_{md}(\textbf{p}_{i}-\textbf{p}_{j})^2]\frac{\rho_{ij}}{\rho_0}.
\end{aligned}
\end{equation}
In this work, the parameter sets which yield a soft (hard) and momentum dependent equation of state with the incompressibility K$_0$=200 MeV (K$_0$=380 MeV) are considered. From now on we refer to the soft and hard EoS as SM and HM respectively. Even though  K$_0$ has been constrained to a relatively narrow range~\cite{Wang:2018hsw,Hartnack:2005tr,Feng:2011dp,Fevre:2015fza,Danie02}), SM and HM are still considered in this work to generate data with large differences. Further, although we know the in-medium nucleon-nucleon elastic cross section ($\sigma_{NN}$) is suppressed when compared to the free one, the degree of this suppression is still not completely pinned down \cite{Li:2018bus,Li:2018wpv}. We use the FU3FP1 parametrization of $\sigma_{NN}$ as in our previous works \cite{Li:2018bus,Li:2018wpv}. We also consider  the free $\sigma_{NN}$ in this study. All together, four parameter sets of the UrQMD model listed in Table~\ref{tab1} are used. Their influence on five observable quantities, the nuclear stopping power ($vartl$) from free protons, the directed flow $v_1$=$\langle \frac{p_{x}}{p_{t}} \rangle$, the elliptic flow $v_{2}=\langle\frac{p_{x}^{2}-p_{y}^{2}}{p_{x}^{2}+p _{y}^{2}}\rangle $, yield of free protons and multiplicity for central (0$\leq$$b$$\leq$2 fm) collisions obtained with different model parameter sets are listed in Table~\ref{tab2}. Clearly, observables are affected by model parameters.
 For example, $v_1$ increases by 70\% if the flow obtained from SM-I is compared to that from HM-F. The isospin-dependent minimum span tree (iso-MST) algorithm is used in UrQMD
  model to recognize clusters. The yields of free protons and clusters are very sensitive to cluster recognition parameters (i.e., the maximum distance and relative momentum between two nucleons). To consider this issue, calculations with parameter sets different from the nominal ones (see caption for details) are also presented as SM-I(MST)*. As listed in Table~\ref{tab2}, the number of free protons and M$_{ch}$ also varies a lot with the cluster recognition parameters.

\begin{table}[h]
\centering
\caption{\label{tab1} Four parameter sets of the UrQMD model with different mean-field potential and nucleon-nucleon elastic cross section. }
\setlength{\tabcolsep}{1.5mm}
\begin{tabular}{ccc}
\hline

EoS         &cross section &  mode   \\ \hline
SM          &   free       &  SM-F    \\
SM          &in-medium     &  SM-I     \\
HM          &   free       &  HM-F        \\
HM          &in-medium     &  HM-I       \\ \hline
\end{tabular}
\end{table}

\begin{table}[t]
\centering
\caption{\label{tab2} Observable quantities (i.e., $v_1$ slope and $v_2$ of free protons at midrapidiy, the yield and $vartl$ of free protons, the total charged multiplicity M$_{ch}$) obtained with different UrQMD parameter sets. We note here that the results listed in this table cannot be compared directly with experimental data, because events with flat $b$-dependent are simulated. To compare with experimental results, calculations with $b$-weighted events should be used.}
\begin{threeparttable}
\footnotesize
\setlength{\tabcolsep}{1.5mm}
{
\begin{tabular}{c|ccccc|} \hline
mode &$v_1$ slope &$v_2$  &yield  & $vartl$ & $M_{ch}$\\ \hline
SM-F                     & 0.14                                                                   & -0.0046$\pm$0.0011                                               & 44.98       & 0.94    & 87.60    \\
SM-I                     & 0.11                                                                   & -0.0024$\pm$0.0012                                                   & 43.97    & 0.91    & 86.35    \\
SM-I(MST)$^*$                & 0.097                                                                  & -0.0043$\pm$0.0013                                                   & 36.95 & 0.92    & 81.84       \\
HM-F                     & 0.19                                                                   & -0.0077$\pm$0.0010                                                   & 50.07 & 0.97    & 90.51  \\
HM-I                     & 0.15                                                                   & -0.0043$\pm$0.0010                                                   & 49.14 & 0.89    & 89.24    \\ \hline
\end{tabular}
}
\begin{tablenotes}
       \footnotesize
       \item[*] This is SM-I mode in combination with MST algorithm (two nucleons with relative distance $\Delta r$ $\leq$ 4.8 fm and relative momentum $\Delta p$$\leq$0.25 GeV/c are considered to belong the same cluster) to recognition fragments.
       While in other cases, the isospin dependent MST algorithm with $\Delta r^{pp}$$\leq$2.8 fm, $\Delta r^{nn}$$\leq$3.8 fm,
       $\Delta r^{np}$$\leq$3.8 fm, and $\Delta p$$\leq$0.25 Gev/c is used.
     \end{tablenotes}
    \end{threeparttable}
\end{table}

For each parameter set, 60 000 events of $^{132}$Sn+$^{124}$Sn collisions with a uniform impact parameter distribution in 0$\leq$$b$$\leq$7 fm at 270 MeV$/$nucleon are simulated. Data obtained from 50 000 of these events are classified as the training data while the remaining 10,000 events are the testing data. Normally, the $bdb$ weighted distribution (i.e., the number of events with an impact parameter $b$ being proportional to $b$) due to the collision geometry is used in transport model simulations. In Refs.~\cite{SBASS,PLBZK,JPGLFP}, large bias between predicted and true impact parameter has been observed for the central collisions because a very small fraction of the total events is central collisions \footnote{Another possible reason is that physical fluctuation is larger in central collisions than in peripheral ones.} \cite{PRCC,NPAA,PB}. To avoid this issue, events with flat distribution of impact parameter are simulated.

Usually, the transverse momentum $p_{t}$= $\sqrt{p_x^2+p_y^2}$ and rapidity $y_{z}$=$\frac{1}{2}\ln[\frac{E+p_{z}}{E-p_{z}}]$ of charged particles\footnote{Throughout this manuscript, transverse momentum per nucleon is used instead of transverse momentum for clusters.} can be measured in heavy-ion experiments. In Ref.~\cite{fopi}, the reduced rapidity $y_{0}$=$y_z$/$y_{pro}$ is used instead of $y_z$. Here, $y_{pro}$ denotes the rapidity of the projectile in the c.o.m system. In order to minimize preprocessing, the two-dimensional $p_{t}$ and $y_{0}$ spectra of all charged particles with 30$\times$30 grid is also used as the input dataset. $p_{t}$ ranges from 0 to 1 GeV/c and $y_{0}$ ranges from -2 to 2. This two-dimensional $p_{t}$ and $y_{0}$ spectra of all charged particles is labelled as \textit{DATASET1}.

For \textit{DATASET2}, we use 7 input features or observables obtained from $^{132}$Sn+$^{124}$Sn at 270 MeV$/$nucleon.

Five of the seven features are the number of deuteron, triton, and helium isotopes $N(D,T,He)$, the averaged transverse momentum of deuteron, triton, and helium isotopes $N(D,T,He)$$p_{t}$,
the number of free protons at mid-rapidity ($|y_{0}|$$\leq$0.5) $Np$, the averaged transverse momentum of free protons at mid-rapidity $N$$p_{t}$. The remaining two features are: $ERAT$ for free protons, defined as
\begin{equation}\label{eq3}
ERAT=\frac{\sum_{i}[p_{t i}^{2}/(2m+E_{i})]}{\sum_{i}[p_{z i}^{2}/(2m+E_{i})]},
\end{equation}
 and the transverse kinetic energy E$_{\perp}$ for light charged particles with the charge number Z=1 and Z=2, defined as
 \begin{equation}\label{eq4}
  E_{\perp}=\sum_{Z=1,2}\frac{p_{t}^2}{2m}.
  \end{equation}

 We use three different ML algorithms with dataset as listed in Table.~\ref{tab3}.
 To assess the accuracy of the reconstruction of the impact parameter, the performance of different algorithms can be quantified by the mean absolute error:
  \begin{equation}\label{3}
  \Delta b = \frac{1}{N_{event}}\sum_{i=1}^{N_{events}}|b_{i}^{true}-b_{i}^{pred}|.
  \end{equation}
 Here, $b_i^{true}$ is the true impact parameter of each event and $b_i^{pred}$ is the predicted one from different algorithms.

\begin{table}[h]
\centering
\caption{\label{tab3} Four different ML algorithms with dataset. }
\begin{threeparttable}
\setlength{\tabcolsep}{1.5mm}
\begin{tabular}{ccc}
\hline
Algorithms                 &   dataset                           &label       \\ \hline
CNN                        &   \textit{DATASET1}           &CNNa        \\
LightGBM                   &   \textit{DATASET1}                  &LightGBMa        \\
ANN$^{1}$                  &    \textit{DATASET2}              &ANNb    \\
LightGBM                   &    \textit{DATASET2}                       &LightGBMb       \\\hline

\end{tabular}
\begin{tablenotes}
       \footnotesize
       \item[1]ANN is more suitable for data with 7 input features than CNN. See details in Section \ref{section3}.
     \end{tablenotes}
    \end{threeparttable}
\end{table}

\section{ANN, CNN and LightGBM algorithms}\label{section3}
ANN composes of multiple dense layers which are connected to each other with many non-linear functions~\cite{BPE,NN}.
When ANN solves a problem, it converts the input data into a one-dimensional vector before the model is trained. More ANN parameters are needed to handle input data with larger dimension.
For image data, ANN easily losses its spatial characteristics resulting in unsatisfactory training results.
Hence ANN is more suitable for tabular data like \textit{DATASET1}. To solve the problem of information loss and too many parameters when ANN extracts spatial  features, new neural networks have been developed, such as CNN \cite{CNN1,CNN2}.

CNN algorithm is one of feed-forward neural networks that includes convolution calculations and has a deep structure.
It is one of the representative algorithms of deep learning at present. It can perform shift-invariant classification of input information, so it is also called the shift-invariant artificial neural network (SIANN).
Several important concepts such as local receptive fields and shared weights are employed in CNN, so it has much less parameters compared with ANN.
Usually, batch normalization, dropout and activation are added between the convolutional layer and the pooling layer to prevent gradient dispersion, overfitting and underfitting~\cite{BN,DP,AC}.

Our CNN architecture includes two convolutional layers, each layer has 64 filters of $5\times5$ size following batch normalization, PReLU activation and dropout with rate 0.3.
In order to filter the features again, we add an averaged pooling layer which is $2\times2$ size scanning through the data from the previous layer.
Finally, in the fully-connected layer, the feature maps from the averaged pooling layer will be flattened and a dropout with a rate of 0.5 is added.
With the suitable adjustment to these parameters, a more stable and reliable CNN version can be prepared for further use.

Decision tree, which is another class of ML algorithm, is more interpretable and is commonly used in the study of physical problems \cite{XIA,BPR,HJY}.
We use LightGBM in this work, which is a decision tree$-$based algorithm with the following advantages~\cite{LGB}: (1) Faster training efficiency, (2) Low memory usage, (3) Higher accuracy, and (4) Tackling large-scale data.
The core of the LightGBM is the histogram-based algorithm, Gradient-based One-Side Sampling (GOSS) and Exclusive Feature Bundling (EFB).
The histogram-based algorithm buckets continuous feature values into discrete bins.
After counting the features of all data, it will find the optimal split value according to the gain of each feature.
Then, these divided features are combined to obtain the contribution value of a tree $f_{k}(x_{i})$, where $x_{i}$ represents the features of each sample and $k$ is a certain decision tree.
Each $f_{k}$ corresponds to an independent tree structure with leaf weights. For a given sample, we  calculate the final prediction $y_{i}$ by summing up the $f_{k}$,
$ y_{i} =  \sum_{k=1}^K f_{k}(x_{i})$. Here $K$ is the total number of decision trees.
Not only do GOSS and EFB avoid information loss and improve computing efficiency, but also solve the problem of sparse feature caused by high-dimensional data.
With these advantages, LightGBM stands out in many data science applications. For more details, readers can consult its official documentation~\cite{LightGBM}.


\section{Results and Discussions}\label{section4}
\subsection{Reconstruction results of \textit{DATASET1} and \textit{DATASET2}}\label{subsection2}
\begin{figure}[htbp]
\begin{centering}
\includegraphics[width=0.4\textwidth]{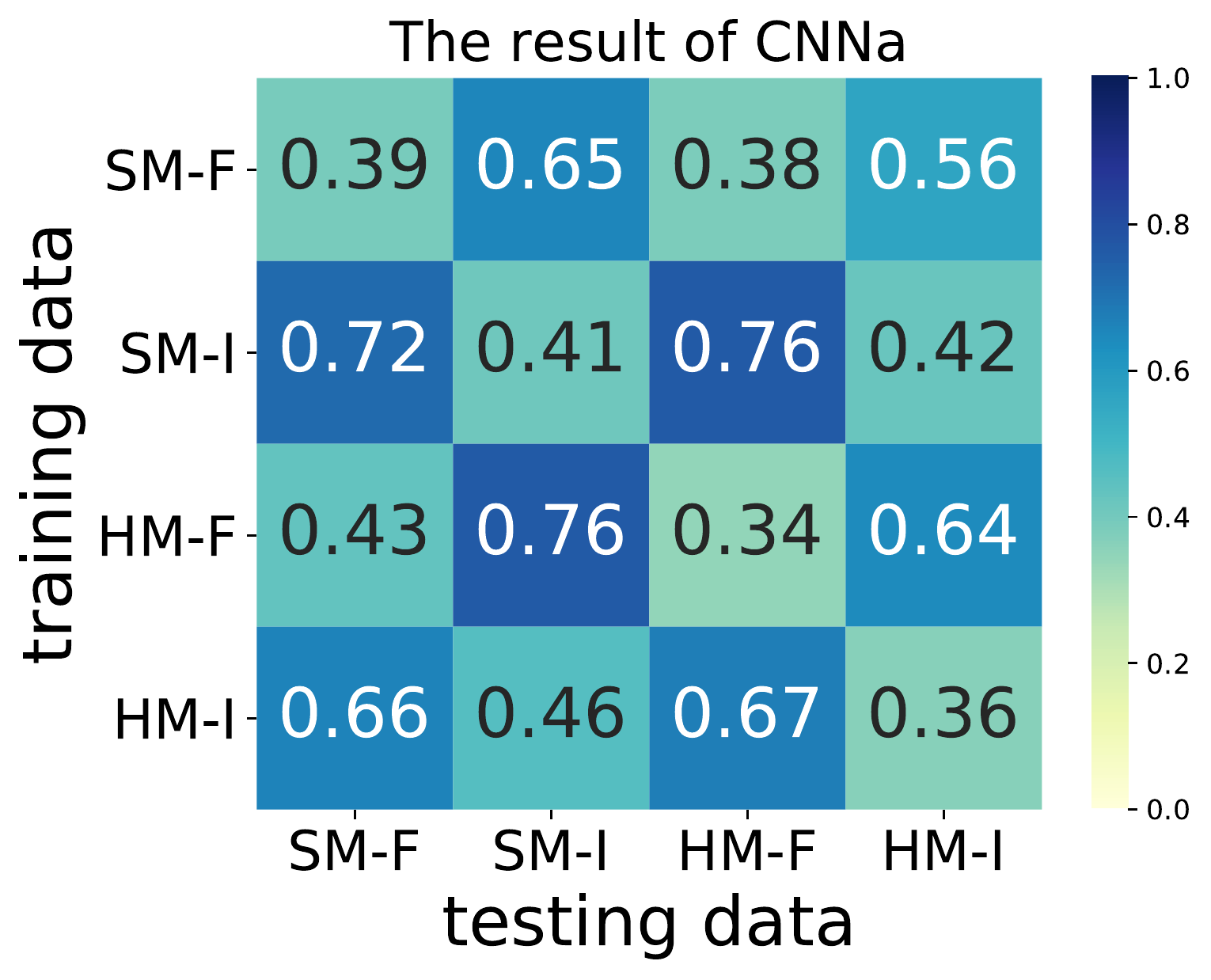}
\includegraphics[width=0.4\textwidth]{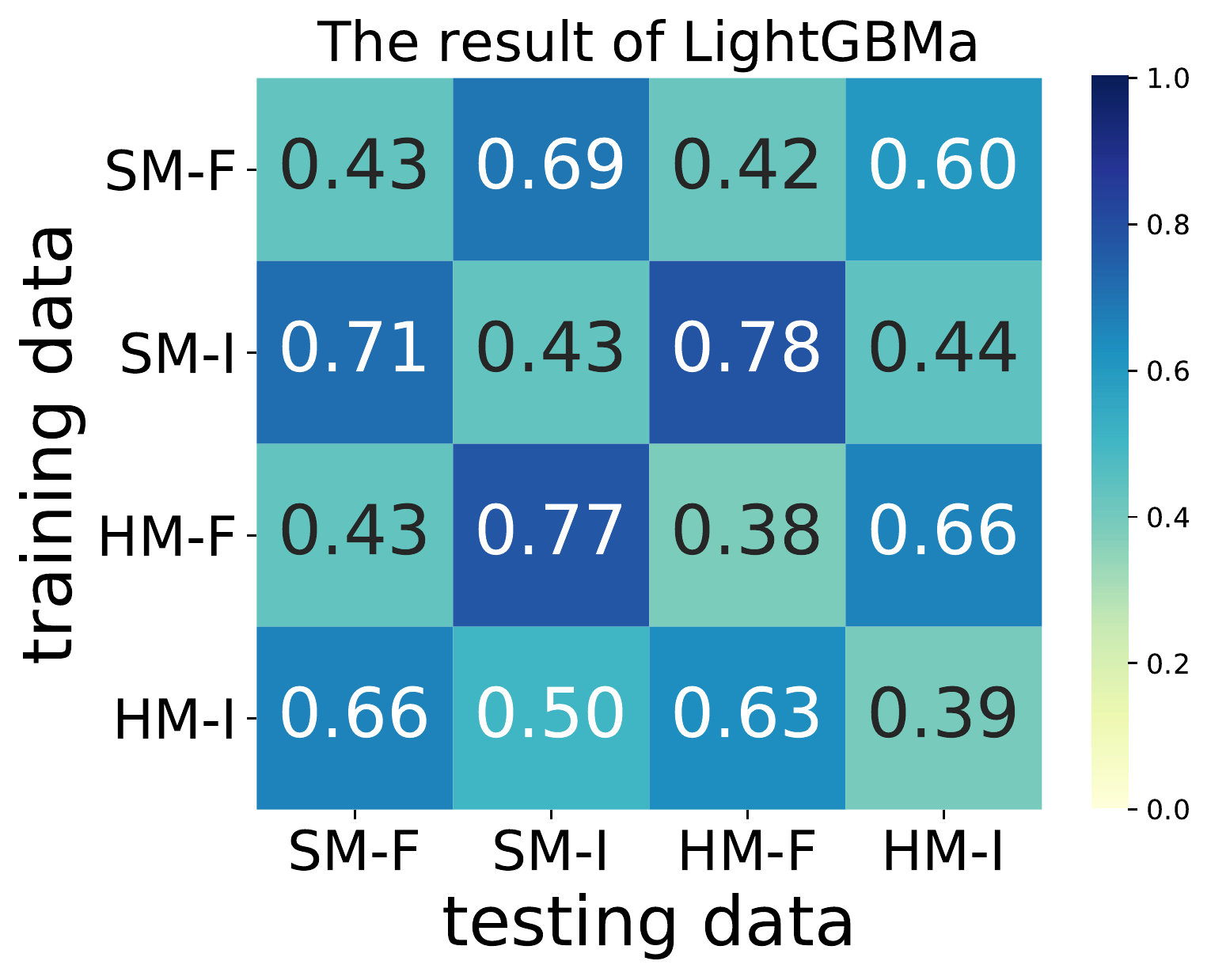}
   \caption{The results of the CNNa and LightGBMa algorithms when \textit{DATASET1} is used. The number in each cell denotes $\Delta b$ for the testing data (generated with the vertical labelled mode) by using the training data (generated with the horizontal labelled mode). The statistical error due to the randomness in the testing data was estimated to be smaller than 1\% by comparing parallel testing data, being therefore negligible.}
   \label{fig3}
   \end{centering}
\end{figure}

The results of CNNa and LightGBMa in which \textit{DATASET1} serves as the input training data are displayed in Fig.~\ref{fig3}. As can be seen, $\Delta b$ is about 0.3-0.4 fm (numbers along the diagonal) if both the training data and testing data are generated from the same UrQMD model parameter set. By using training and testing data obtained from different parameter sets (off diagonal),  e.g., the two-dimensional $p_{t}$ and $y_{0}$ spectra of all charged particles generated with SM-I serves as the training data while simulation data generated with HM-F serves as the testing data, $\Delta b$ is increased to about 0.8 fm. This is understandable due to parameters in both the mean-field potential and collision terms (the two of main ingredients of transport model) being different in SM-I and HM-F modes. In addition, it can be found that $\Delta b$ is affected much more by cross section than by $K_0$. For example, $\Delta b$ for testing data obtained from SM-F by using ML algorithms trained with data from SM-I mode is about 0.7 fm, while by using ML algorithms trained with data from HM-F are about 0.4 fm. This is due to the fact that both $b$ and $\sigma_{NN}$ strongly affect the number of collisions and the final observed particle spectra. Thus, the fingerprint of impact parameter on particle spectra is erased to some extent by varying $\sigma_{NN}$. Furthermore, $\Delta b$ in most cases obtained with CNNa is slightly smaller than that obtained with LightGBMa, indicating that CNN has a better performance on \textit{DATASET1} than LightGBM. However, considering the fact that LightGBM is at least 10 times faster than CNN and does not require a GPU, LightGBM is a better choice for all practical purposes.

\begin{figure}[htbp]
\begin{centering}
\includegraphics[width=0.4\textwidth]{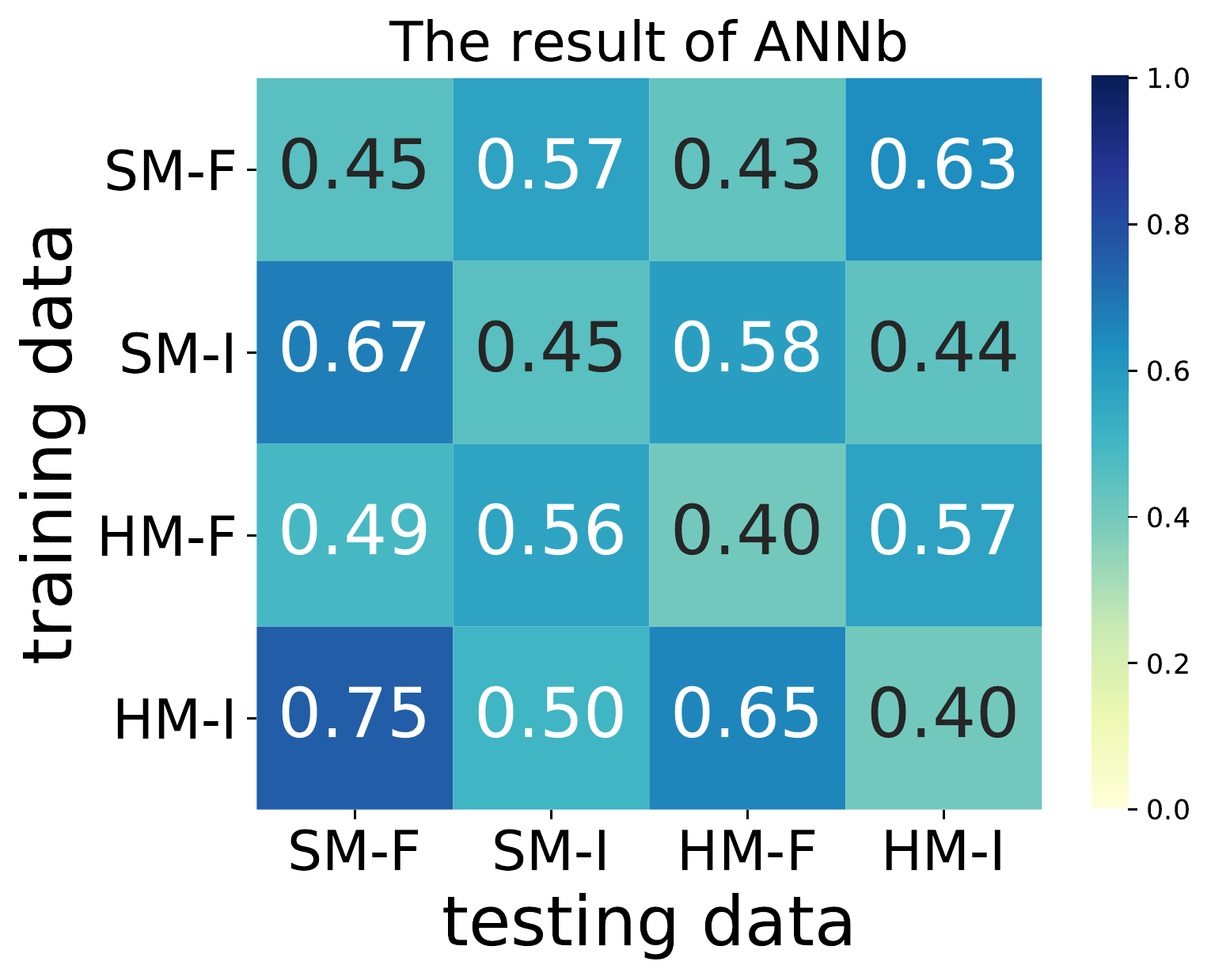}
\includegraphics[width=0.4\textwidth]{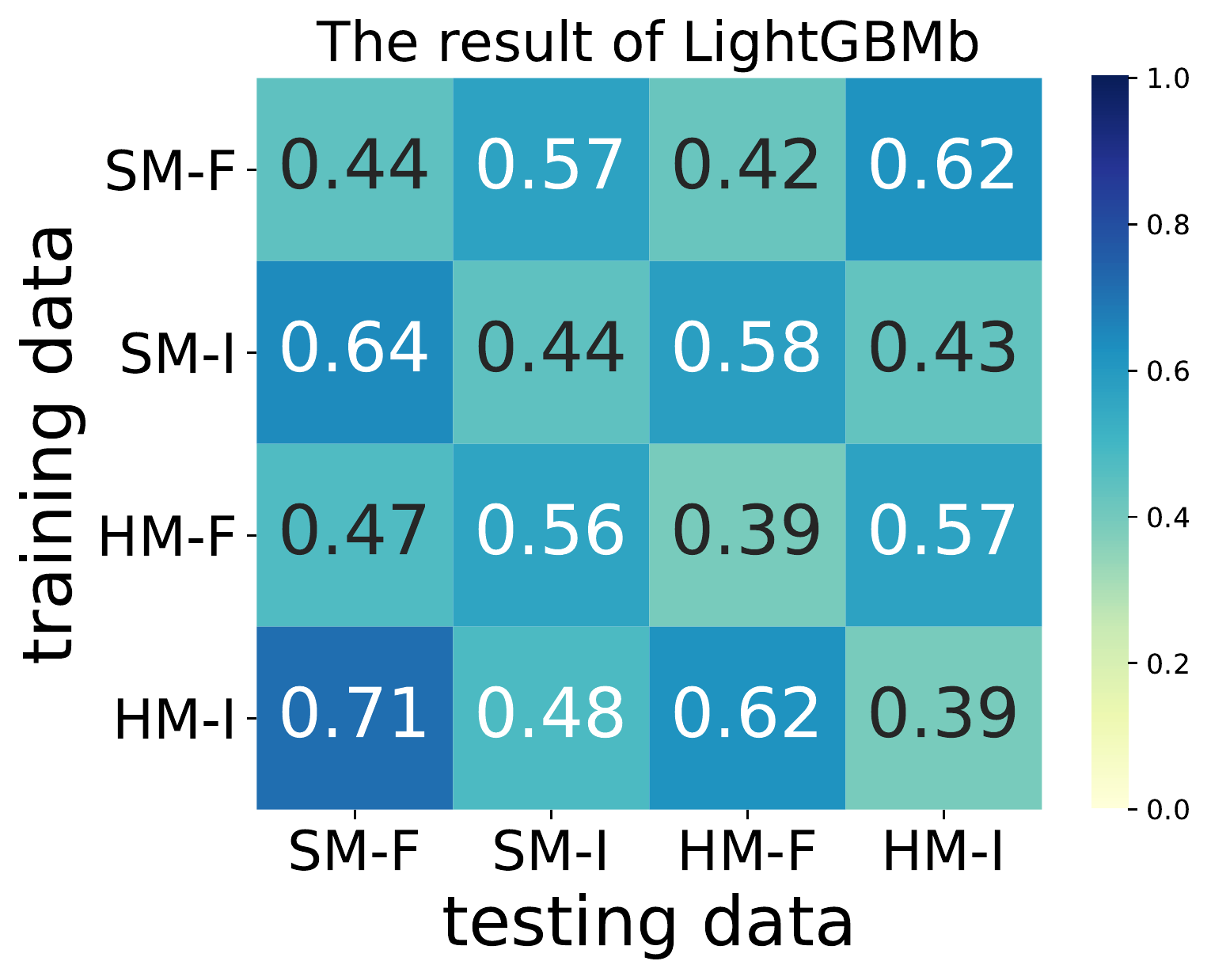}
   \caption{The results of the ANNb and LightGBMb algorithms by using \textit{DATASET2}.
The number in each cell denotes $\Delta b$ for the testing data (generated with the vertical labelled mode) by using the training data (generated with the horizontal labelled mode). }
   \label{fig4}
      \end{centering}
\end{figure}

Fig.~\ref{fig4} shows $\Delta b$ obtained with ANNb and LightGBMb algorithms by using \textit{DATASET2}. For both training data and testing data generated from the same parameter set, $\Delta b$ are about 0.4-0.45 fm which is slightly larger than the corresponding values displayed in Fig.~\ref{fig3}. We observe the same trend that the diagonal numbers are smaller than off diagonal.  In addition, $\Delta b$ for training and testing data generated from parameter sets with the same mean-field potential, but different $\sigma_{NN}$ are also larger than other cases, indicating again $\sigma_{NN}$ has a stronger effect than the mean-field potential.
However, even for training data and testing data generated from these different parameter sets, $\Delta b$ is still smaller than 0.8 fm obtained from CNNa even in the worst case.

Regrading the influence of cluster algorithm, by using LightGBMb algorithm trained with data from SM-F, SM-I, HM-F, and HM-I, the $\Delta b$ are 0.72, 0.50, 0.70, and 0.51 fm for testing data obtained with SM-I(MST), respectively. The observation is the same as discussed above, which is simulation using FU3FP1 parametrization for $\sigma_{NN}$ gives better $\Delta b$.  Overall, $\Delta b$ is smaller than 0.8 fm regardless of the model parameters or cluster recognition algorithms sets used to generate data.

\subsection{Impact parameter dependence}\label{subsection3}

\begin{figure}[htbp]
\begin{centering}
\includegraphics[width=0.48\textwidth]{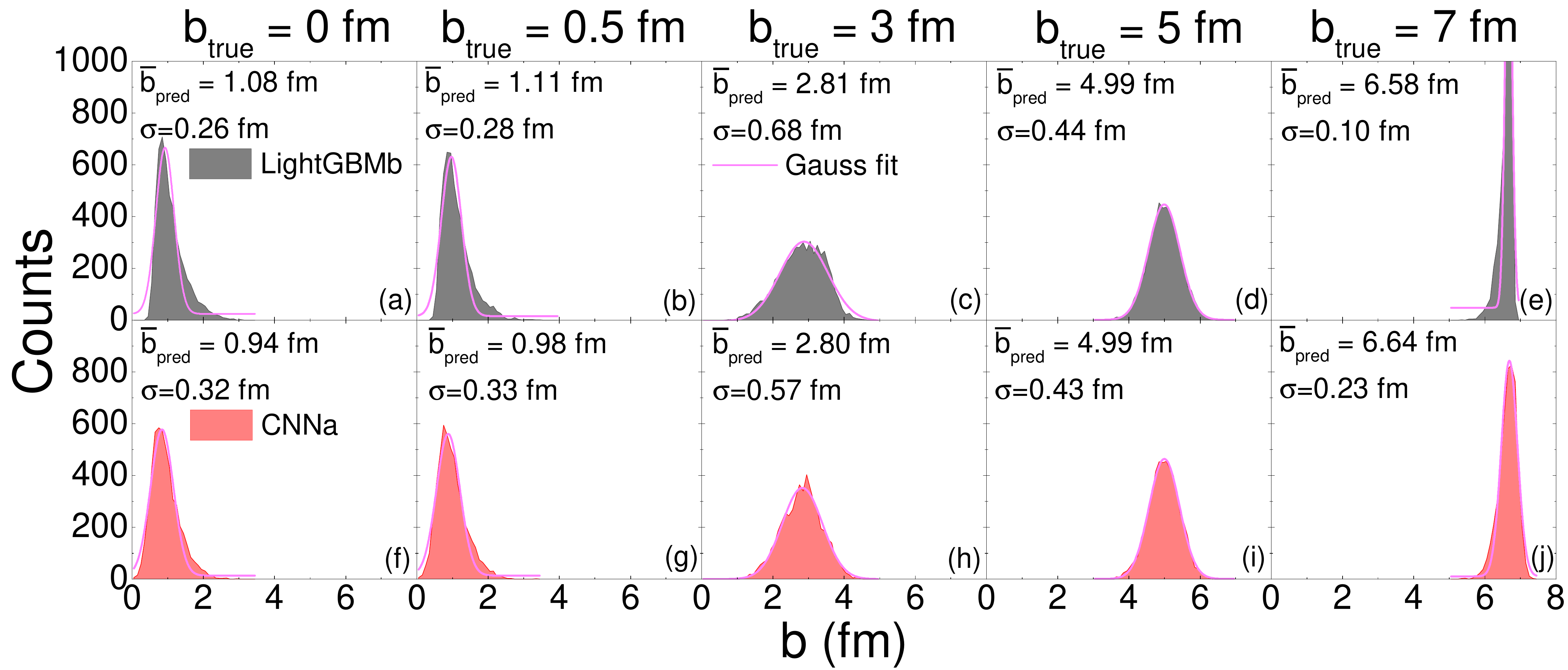}
\caption {\label{fig5}The distribution of the predicted impact parameter from LightGBMb and CNNa algorithms. Both the training data and testing data are generated with SM-I mode. 5 000 $^{132}$Sn+$^{124}$Sn collision events for each impact parameter (from left panel to right, $b$=0, 0.5, 3, 5, and 7 fm) are tested. The pink lines represent the Gaussian fitting ($y=y_{0}+\frac{A}{\sigma\sqrt{2\pi}}exp^-{\frac{(x-\mu)^2}{2\sigma^2}}$) of the distribution. $\bar{b}_{pred}$ and $\sigma$ represent the averaged value of the predicted $b$ and its standard deviation, respectively. }
\end{centering}
\end{figure}

It is observed that $\Delta b$ depends on the impact parameter, and $\Delta b$ is larger in central collisions~\cite{SBASS,PLBZK,JPGLFP}. Fig.~\ref{fig5} shows the distributions of the predicted impact parameter obtained with LightGBMb (top panels) and CNNa (Bottom panels) algorithms. Above 1 fm, the averaged value of $\bar{b}_{pred}$ is close to the true value. For $b_{true}$=0 and 0.5 fm, $\bar{b}_{pred}$ are about 1.0 fm, much larger deviations from $b_{true}$. The random nucleon-nucleon collision processes are much more abundant when $b$ is small, therefore fingerprint of impact parameter on various observables might be washed out by the stochastic process. If the outcomes of collisions with $b$=0 and 1 fm are naturally indistinguishable, but collisions with $b$$\textgreater$1 fm are distinguishable, the $b_{pred}$ for events with $b_{true}$$\textless$1 fm given by the ML algorithm would close to 1 fm in order to get the smallest global loss, because $b_{true}$ varies from 0 to 7 fm. When $\bar{b}_{pred}$ obtained from LightGBMb and CNNa are compared, the latter performs much better in the most central collisions.

\subsection{Explanation of LightGBM algorithm }\label{subsection4}

\begin{figure}[htbp]
\begin{centering}
\includegraphics[width=0.48\textwidth]{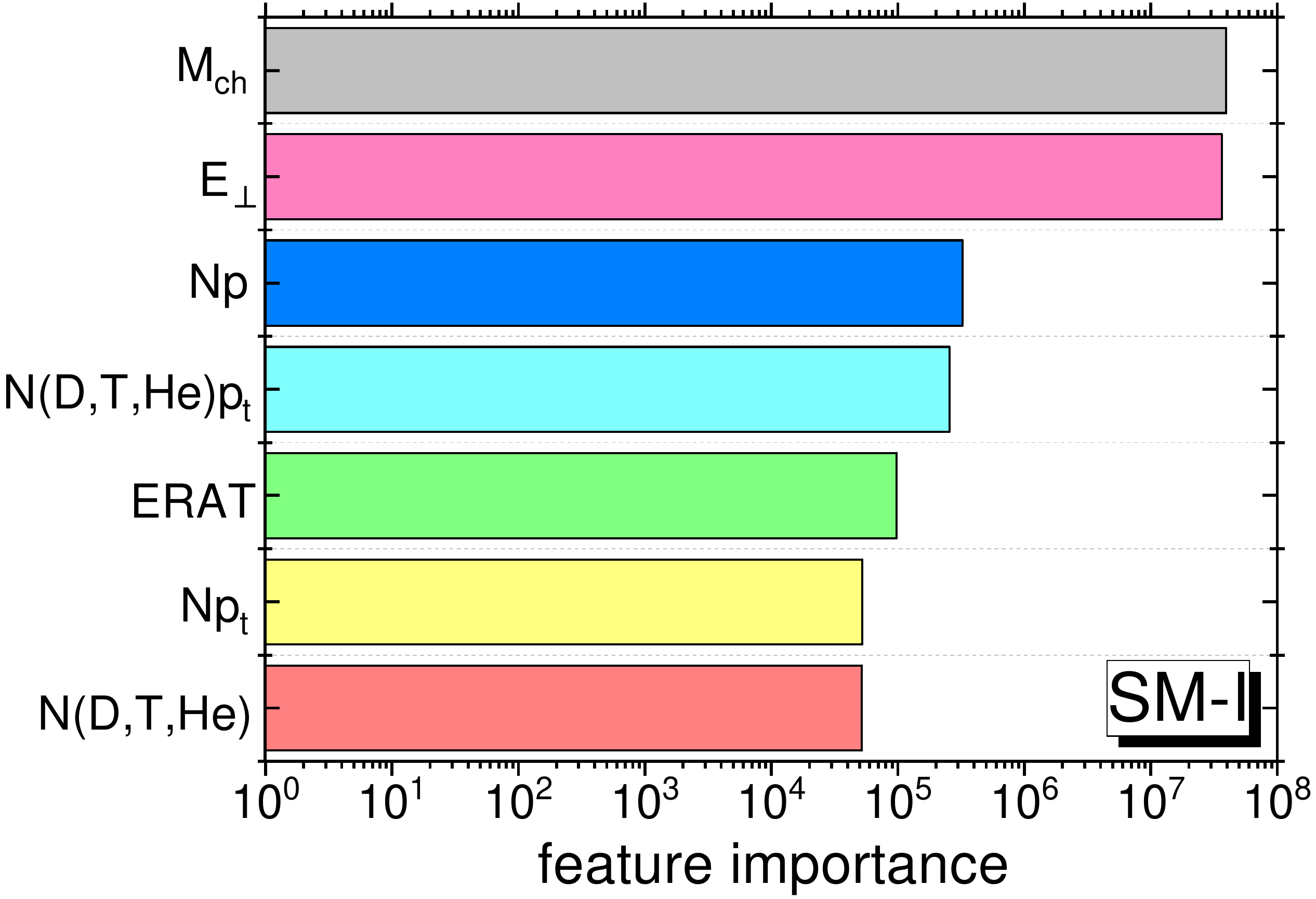}
\caption {\label{fig6} The importance of the 7 features obtained with the \emph{Feature importance} technology of LightGBM algorithm. }
\end{centering}
\end{figure}

\begin{figure}[htbp]
\begin{centering}
\includegraphics[width=0.48\textwidth]{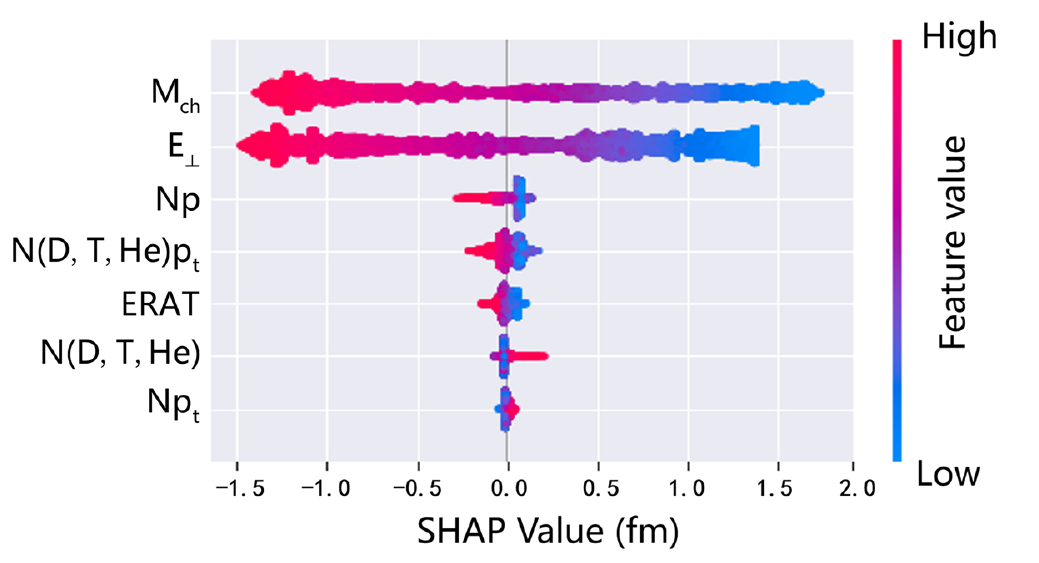}
\caption {\label{fig7} Importance ranking for the 7 input features obtained with SHAP package. Each row represents a feature, and the x-axis is the SHAP value which shows how important a feature is for a particular prediction. Each point represents a sample, and the color represents feature value (red is high, blue is low).}
\end{centering}
\end{figure}

\begin{figure}[htbp]
\begin{centering}
\includegraphics[width=0.48\textwidth]{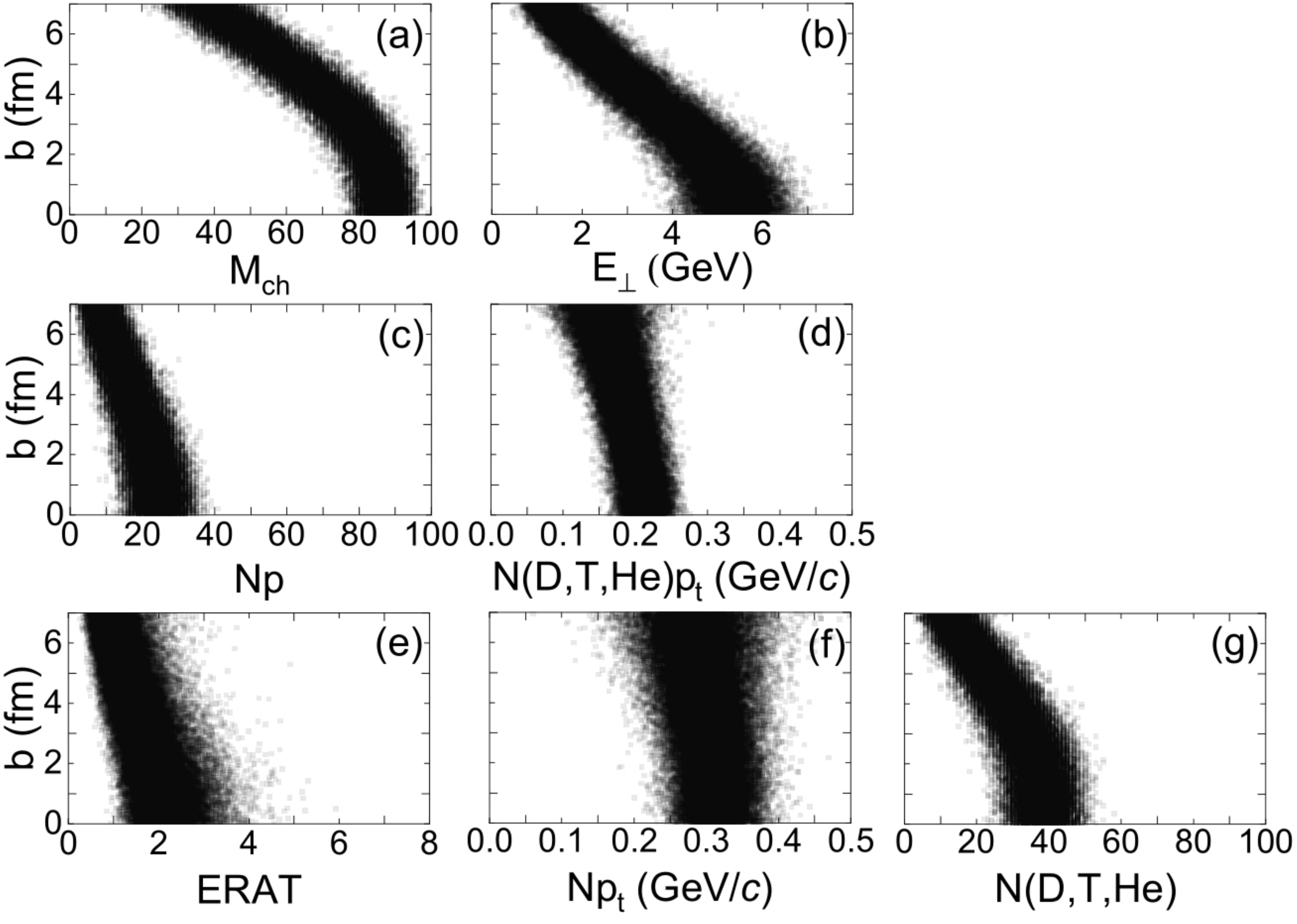}
\caption {\label{fig8}The correlation between the impact parameter and the 7 input features.}
\end{centering}
\end{figure}

\begin{figure}[htbp]
\begin{centering}
\includegraphics[width=0.48\textwidth]{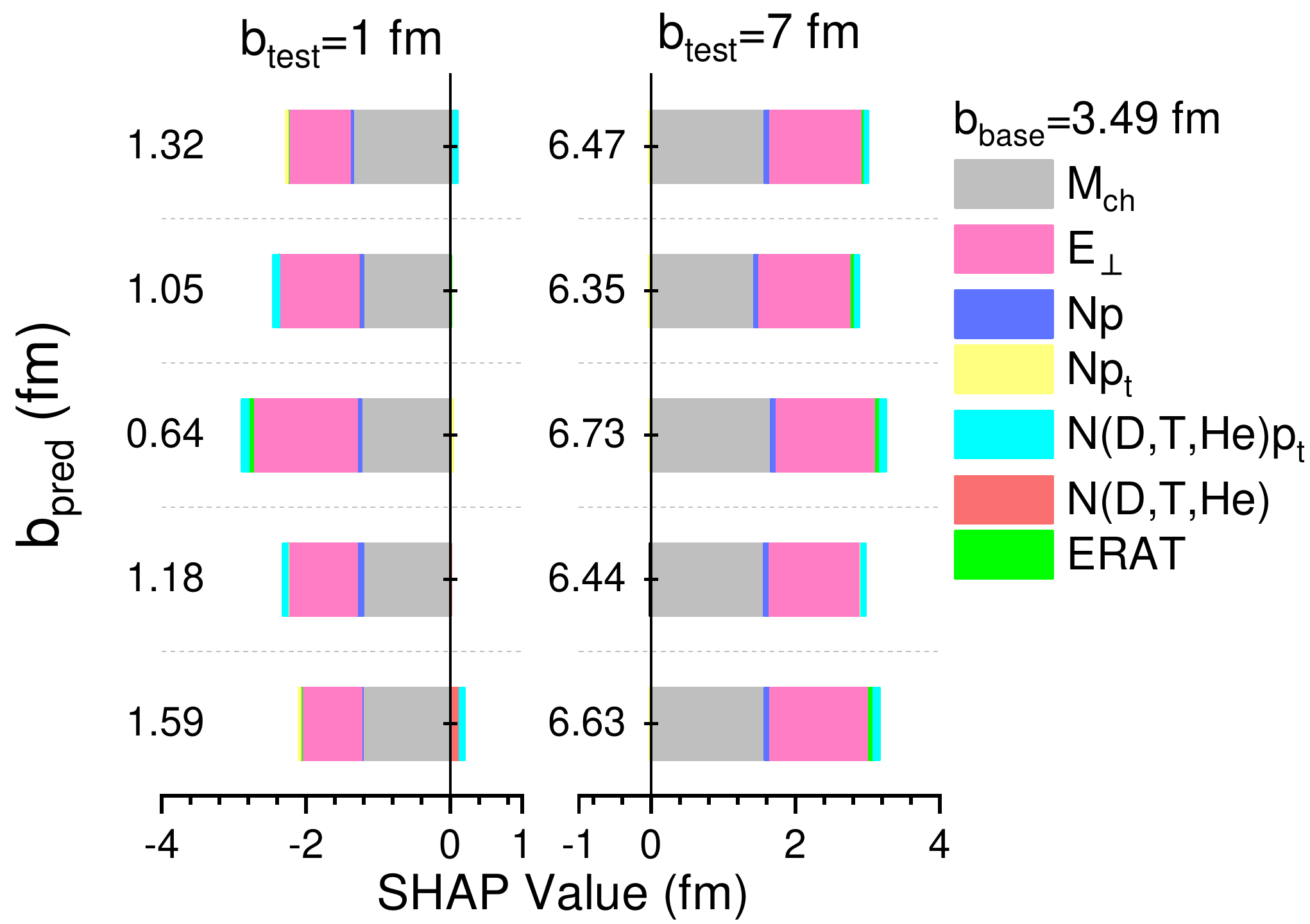}
\caption {\label{fig10} The contribution of each feature to a prediction ($b_{pred}$) obtained with the SHAP algorithm, which pushes the prediction of the model from the base value to the final value (model output $b_{pred}$). The base value is the mean value of the model predicted value on the training data, here $b_{base}$ =3.49 fm. Results from 5 random events for each tested impact parameter ($b_{test}$=1 and 7 fm) are illustrated as examples. }
\end{centering}
\end{figure}

\begin{table}[h]
\centering
\caption{\label{tab5} The Pearson correlation coefficient among the 7 features and the impact parameter.}
\setlength{\tabcolsep}{1mm}
\begin{tabular}{c|ccccccc}
\hline

       &E$_{\perp}$   &   M$_{ch}$    &N(D,T,He)         &Np   &N(D,T,He)p$_{t}$  &ERAT    &Np$_{t}$    \\ \hline
b      &    0.94      &0.93           &0.86             &0.82  &0.68                 &0.67   &0.29  \\ \hline
\end{tabular}
\end{table}
 LightGBM is very explainable whereas CNN is often treated as a black box. Explainable ML algorithms are usually preferred, especially when they are applied to solve physical problems~\cite{EX1,EX2} because understanding what happens when ML algorithms make predictions could help us make better use of the outputs. To understand how the LightGBM algorithm gives a particular result and to develop insight into what ML algorithm has learned, \emph{Feature importance} technology of LightGBM and SHapley Additive exPlanation (SHAP)~\cite{SHAP} are applied to show which features have the greatest effect on the determination of impact parameter.

 Fig.~\ref{fig6} and Fig.~\ref{fig7} illustrate the ranking of importance of the 7 features. In both figures, M$_{ch}$ and E$_{\perp}$ are ranked as the two most important features, while the importance of the other five features are similar and very weak. To understand the feature importance, the correlation between the impact parameter and the 7 input features are plotted in Fig.~\ref{fig8}. The impact parameter $b$ are much more correlated with M$_{ch}$ and E$_{\perp}$ than the others, which implies that they can serve as good candidates for determining $b$. In addition, we calculate their Pearson correlation coefficient (PCC). PCC is often used to measure the linear correlation between two variables in statistics. PCC between $b$ and M$_{ch}$, as well as $b$ and E$_{\perp}$ are close to 1, implying a strong linear correlation between them. Meanwhile, one may find some inconsistencies in Fig.~\ref{fig6}$-$Fig.~\ref{fig8} and Table~\ref{tab5}. For example, PCC between $b$ and $N(D,T,He)$ is the third largest one, but $N(D,T,He)$ ranks as almost the most irrelevant feature in Fig.~\ref{fig6} and Fig.~\ref{fig7}. This could occur if the ML algorithm learns not only the linear but also non-linear multifaceted relationship between the output and the input features.

  SHAP is a model interpretation package developed by Python that interprets the output of ML model. For each test sample, SHAP value can be calculated with $ b_{i} = b_{base} + f(x_{i1})+f(x_{i2})+\ldots+f(x_{ik})$, where $f(x_{ik})$ is the SHAP value of $x_{ik}$. Here $x_{ik}$ represents the value of the $k-$th input feature of the $i-$th sample, $b_{i}$ is the predicted $b$ for the $i-$th sample and $b_{base}$ is the mean value of the all samples. In the present work, as the output $b$ is uniformly distributed from 0 to 7 fm, thus the $b_{base}$ is about 3.49 fm. Fig.~\ref{fig10} displays the contribution of each feature to a certain prediction. The results for random five samples for each impact parameter ($b$=1 and 7 fm) are displayed. When the impact parameter is less (greater) than $b_{base}$, the SHAP value of each feature is basically negative (positive). As observed in Fig. \ref{fig8}, for a smaller $b$, both values of M$_{ch}$ and E$_{\perp}$ are larger. It is understandable as more particles and transverse energy may produced from the more central collision. This is also the reason why the SHAP value of the red dots (samples with high values of M$_{ch}$ and E$_{\perp}$) in the first two rows are more negative in Fig.~\ref{fig7}.

  Overall, it can be found that M$_{ch}$ and E$_{\perp}$ are the two most important input features for reconstructing the impact parameter while $N(D,T,He)$ and $N$$p_{t}$ are listed as being the most irrelevant features. Based on the ranking importance, we can reduce the number of features by taking a subset of the most important features. We have checked that, the performance does not change if $N(D,T,He)$ and $N$$p_{t}$ are not included as features in the training.

\section{Summary and outlook}\label{section5}
In this work, three popular ML algorithms, ANN, CNN and LightGBM, are applied to determine the impact parameter by using either the proton spectra or 7 features generated with the UrQMD model. To test the generalizability of the trained ML algorithms, four different UrQMD model parameter sets are applied to generate the data. It is found that the mean absolute error between the true impact parameter and the estimated one $\Delta b$ can be smaller than 0.45 fm if training and test sets are taken from the UrQMD model with the same parameter set, while $\Delta b$ increases to 0.8 fm if the training and testing data are taken from different parameter sets in the UrQMD model. Furthermore, the feature importance is obtained with LightGBM algorithm based on \textit{Feature importance} technology and SHAP. The total number of charged particles M$_{ch}$ and the transverse kinetic energy E$_{\perp}$ for light charged particles are the two most relevant features for determining the impact parameter, and this can be understood from the distribution of impact parameter as functions of M$_{ch}$ and E$_{\perp}$.

The generalizability of the trained ML algorithms is tested by using training and testing data generated from different model parameter sets. Although observables are strongly affected by the model parameters, the extracted $\Delta b$ is still smaller than 0.8 fm, implying the trained ML algorithms are robust approaches. This gives us confidence that the trained ML algorithms with data generated by theoretical models can be applied to determine the impact parameter in real experimental data as shown in Ref.~\cite{tommy}.


\section*{Acknowledgements}

The authors acknowledge
support by the computing server C3S2 in Huzhou University. The work is supported in part by the National Natural Science Foundation of China Nos. U2032145, 11875125, and 12047568,
and the National Key Research and Development Program of China under Grant No. 2020YFE0202002, and the ``Ten Thousand Talent Program" of Zhejiang province (No. 2018R52017). This work was partly supported by the U.S. Department of Energy, USA under Grant Nos. DE-SC0021235, DE-NA0003908, and U.S. National Science Foundation Grant No. PHY-1565546.

\end{document}